%% file: main.tex
\documentclass{article}

\usepackage[a4paper, total={6in, 8in}]{geometry}

\usepackage{amsmath}
\usepackage{graphicx}
\usepackage[colorlinks=true, allcolors=blue]{hyperref}

\usepackage[acronyms]{glossaries}

\usepackage{authblk}
\usepackage{orcidlink}
\usepackage{rorlink}

\usepackage{csquotes}
\usepackage[backend=biber, sortcites, defernumbers=true]{biblatex}
\usepackage{subcaption}

\usepackage{tabularx}
\usepackage{rotating}
\usepackage{multirow}
\usepackage{array}
\newcolumntype{C}[1]{>{\centering\arraybackslash}p{#1}}
\newcolumntype{L}[1]{>{\raggedright\arraybackslash}p{#1}}
\newcolumntype{M}[1]{>{\centering\arraybackslash}m{#1}}
\usepackage{pifont}
\newcommand{\cmark}{\ding{51}}%
\newcommand{\xmark}{\ding{55}}%

\title{Towards a Metadata Schema for Energy Research Software}

\begin{document}

\author[1,2]{Stephan Ferenz\orcidlink{0000-0001-9523-7227}\thanks{Corresponding Author: \href{mailto:stephan.ferenz@uol.de}{stephan.ferenz@uol.de}}}
\author[2]{Oliver Werth\orcidlink{0000-0002-6767-5905}}
\author[1,2]{Astrid Nieße\orcidlink{0000-0003-1881-9172}}
\affil[1]{Department for Computer Science, Carl von Ossietzky Universität Oldenburg, Oldenburg, Germany \rorlink{https://ror.org/033n9gh91}}
\affil[2]{Energy Division, OFFIS - Institute for Information Technology, Oldenburg, Germany \rorlink{https://ror.org/003sav189}}
\date{January 14, 2026}

\input{content/acronyms}

\maketitle

\begin{abstract}
Domain-specific metadata schemas are essential to improve the findability and reusability of research software and to follow the FAIR4RS principles. However, many domains, including energy research, lack established metadata schemas. To address this gap, we developed a metadata schema for energy research software based on a requirement analysis and evaluated it through user testing. Our results show that the schema balances the need for formalization and interoperability, while also meeting the specific needs of energy researchers. Meanwhile, the testing showed that a good presentation of the required information is key to enable researchers to create the required metadata. This paper provides insights into the challenges and opportunities of designing a metadata schema for energy research software.
\end{abstract}

\glsunset{FAIR}
\input{content/introduction}
\input{content/methods}

\input{content/results}

\input{content/discussion}

\section*{Acknowledgments}
The authors would like to thank the German Federal Government, the German State Governments, and the Joint Science Conference (GWK) for their funding and support as part of the \href{https://nfdi4energy.uol.de/}{NFDI4Energy} consortium. The work was funded by the German Research Foundation (DFG) – 501865131 within the German National Research Data Infrastructure (NFDI, \url{www.nfdi.de}).

\printbibliography[title=References]

\end{document}

%% file: content/acronyms.tex
\newacronym{FAIR}{FAIR}{Findable, Accessible, Interoperable, Reusable}
\newacronym{CFF}{CFF}{Citation File Format}
\newacronym{URI}{URI}{Unique Resource Identifier}
\newacronym{SHACL}{SHACL}{Shapes Constraint Language}
\newacronym{SMECS}{SMECS}{Software Metadata Extraction and Curation Software}
\newacronym{UTAUT}{UTAUT}{Unified Theory of Acceptance and Use of Technology}
\newacronym{TAM}{TAM}{Technology Acceptance Model}
\newacronym{OEP}{OEP}{Open Energy Platform}
\newacronym{SUS}{SUS}{System Usability Scale}
\newacronym{API}{API}{Application Programming Interface}
\newacronym{openmod}{openmod}{Open Energy Modeling Initiative}

\glsdisablehyper

%% file: content/introduction.tex
\section*{Introduction}
\label{sec:intro}

The \gls{FAIR}4RS (Findable, Accessible, Interoperable, and Reusable for research software) principles~\cite{chuehong_FAIR_2022} provide a crucial framework to enhance the overall reusability of research software. Especially domain-specific metadata schemas have been demonstrated to be essential to improve the findability and reusability of research software~\cite{chuehong_FAIR_2022,barker_Introducing_2022}. While some research domains have already established domain-specific metadata schemas for research software, such as life science \cite{ison_BiotoolsSchema_2021} and geoscience \cite{gil_OntoSoft_2016}, other domains are currently lagging behind. To enable \gls{FAIR} research software in these domains, the development of domain-specific metadata schemas is a critical step.

One of these domains is energy research which focuses on understanding, analyzing, improving, and designing energy systems and components, encompassing social, economic, environmental, and technical aspects. This research heavily relies on research software, such as data analytics and simulation tools for energy components or whole energy systems \cite{ferenz_Requirements_2025}. 
There already exist first approaches for metadata schemas for energy research software, like the factsheets on the \href{https://openenergyplatform.org/}{\gls{OEP}}. These approaches are still limited with respect to formalization, interoperability with general metadata schemas, and a profound requirement analysis as foundation \cite{ferenz_Requirements_2025}. Therefore, the existing approaches are not sufficient enough to support FAIR research software in the energy domain.

As a first step to address these issues towards \gls{FAIR}4RS, we presented a requirement analysis based on 32 interviews with energy researchers in \cite{ferenz_Requirements_2025}. In addition to these requirements, a metadata schema for energy research software should 
fulfill further requirements, such as interoperability with existing schemas and simplicity for users creating metadata based on the schema. In this paper, we address the question how a metadata schema for energy research software should be designed to balance these different requirements. To answer this question, we developed a metadata schema for energy research software (\textit{ERSmeta}) and evaluated it by letting energy researchers create metadata for their own software.

The remaining paper is organized as follows:  First, we present the 'Related Work'. Afterwards, 'Methods' presents the process used to develop the metadata schema and describes the setup for the evaluation. Then, we present our 'Metadata Schema: ERSmeta' and show the outcomes of our 'Evaluation of the Metadata Schema'.  Finally, 'Discussion' gives an overview on the advantages and limitations of our approach and shows further research directions.

%% file: content/methods.tex
\section*{Related Work}
\label{subsec:rw}

In this section, we give an overview on existing metadata schemas for research software based on our preliminary work in \cite{ferenz_Requirements_2025} and \cite{ferenz_Improved_2023}. All presented approaches are summarized in \autoref{Tab:rs}.
First, we examine general approaches and approaches from other domains.

\textit{CodeMeta} \cite{jones_CodeMeta_2023} is a community-driven metadata standard for research software, based on \textit{schema.org}. Various crosswalks to other metadata schemas already exist, e.g., to the metadata of the GitHub \gls{API}. \textit{CodeMeta} contains multiple elements, some focusing on technical details like file size or supported operating systems and others including administrative information like license. The metadata standard does not have mandatory elements. It supports the use of \glspl{URI} for authors,  contributors, and licenses. The content specific metadata are limited to an application category and keywords.

The \textit{\gls{CFF}} \cite{druskat_Citation_2021} provides essential information for properly citing research software. While it covers some relevant metadata information, it does not define a proper metadata schema. Also, \textit{\gls{CFF}} is limited to general information without including any domain-specific elements.

\renewcommand{\arraystretch}{1.4}
\begin{table}[t]
	\centering
	\caption{Overview of metadata schema for research software (based on \cite{ferenz_Improved_2023} and \cite{ferenz_Requirements_2025})}
	\begin{tabular}{c l c l c l}
		\cmark:      & fulfilled         &  (\cmark):    & partly fulfilled & \xmark:     & not fulfilled    \\
	\end{tabular}
	\begin{tabular}{ C{1.8cm} m{5.35cm} C{2.1cm} C{.7cm} C{1.15cm} C{1.15cm}}
		 & & 
		\rotatebox[origin=l]{90}{\parbox[c]{2.35cm}{Domain}} & 
		\rotatebox[origin=l]{90}{\parbox[c]{2.35cm}{Mandatory \\elements}} &  
        \rotatebox[origin=l]{90}{\parbox[c]{2.35cm}{Reuse elements from existing \\approaches}} &  
		\rotatebox[origin=l]{90}{\parbox[c]{2.55cm}{Use of domain \\ontologies as \\value vocabulary}} \\
		\hline
		\hline
		
		\multirow{4}{*}{{\parbox[b]{1.6cm}{\centering Metadata Schemas}}} & 
        \textit{CodeMeta} (\cite{jones_CodeMeta_2023}) & 
        General & \xmark & \cmark & \xmark \\
          &   \textit{\gls{CFF}} \cite{druskat_Citation_2021} & General & \xmark &  \xmark & \xmark \\
           &  \textit{DataDesc} \cite{kuckertz_DataDesc_2024} & General & \cmark &  \cmark & \xmark \\
       & \textit{biotoolsSchema} \cite{ison_BiotoolsSchema_2021} &  Bioinformatics & \cmark & \xmark & \cmark\\
		\hline
		\multirow{2}{*}{{Ontologies}} & \textit{OntoSoft} \cite{gil_OntoSoft_2015} & Geoscience & \cmark & \xmark & \xmark \\
		
		& \textit{Software Desciption Ontology} \cite{garijo_OKGSoft_2019} & General & \xmark & \cmark & \cmark\\

        \hline 
        \multirow{3}{*}{{\parbox[b]{1.6cm}{\centering Not formalized approaches}}} & 
        Catalog of energy co-simulation components \cite{schwarz_Ontological_2019} & Energy & \xmark & \xmark & (\cmark) \\
        
		& \href{https://wiki.openmod-initiative.org/}{\gls{openmod} wiki} & Energy & \xmark & \xmark & (\cmark) \\
		
        & \href{https://openenergyplatform.org/}{\gls{OEP} factsheets on frameworks} & Energy & \cmark & \xmark & (\cmark) \\
		
		
	\end{tabular}
	\centering
	\label{Tab:rs}
\end{table}

\textcite{kuckertz_DataDesc_2024} introduced \textit{DataDesc}, a metadata schema that focuses on interfaces of research software. This schema reuses multiple elements from \textit{CodeMeta} and allows for a detailed description of research software interfaces. Besides keywords and description it does not cover detailed domain-specific aspects.

For geosciences, \textcite{gil_OntoSoft_2015} developed \textit{OntoSoft}, an ontology that describes research software with six categories: identify, understand, execute, do research, get support, and update.
Gariko et al. \cite{garijo_OKGSoft_2019,garijo_Software_2021} expanded this approach by creating the \textit{Software Description Ontology}, which includes additional descriptions for input and output data based on the \textit{Scientific Variables Ontology}. They also aligned their approach with \textit{CodeMeta} and enabled publishing the metadata in an open knowledge graph, including links to additional instances in the semantic web like \textit{wikidata}.

\textcite{ison_BiotoolsSchema_2021} developed the metadata schema \textit{biotoolsXSD} for the software registry bio.tools in life science. This schema is expressed as an XML schema containing 55 elements, of which 10 are mandatory. The use of the \textit{EDAM} ontology as a value vocabulary is required for some elements, such as function, input, and output. The schema also includes software-specific elements, like programming language, license, and operating system, for which the use of an ontology is not required.

Second, we examine approaches from the energy domain. \textcite{schwarz_Ontological_2019} described a catalog of energy co-simulation components. They used a semantic media wiki to collect information on simulators and the Functional Mockup Interface (FMI) to add descriptions on the simulation interfaces. However, the elements of the catalog, which can be used for a metadata schema, are not described in more detail.

The \gls{openmod} includes a list of energy models in \href{https://wiki.openmod-initiative.org/}{their wiki}. However, the metadata schema is not formalized and controlled vocabularies are not used for the elements or values.

The \href{https://openenergyplatform.org/}{\gls{OEP}} includes information on models and frameworks in energy research. The metadata elements are similar to those of the openmod wiki. Their description is limited and the metadata are also not formalized.

All approaches from the energy domain are not formalized as metadata schema. Additionally, the approaches lack descriptions how the elements were chosen and to which extend requirements were included in this process \cite{ferenz_Requirements_2025}.

\section*{Methods}
\label{sec:method}

Within this section, we describe the process which led to our metadata schema \textit{ERSmeta} and its evaluation. First, we present our method used for the development of the metadata schema. Afterwards, we outline our evaluation method.

\subsection*{Development of the Metadata Schema}
\label{subsec:dev_schema}

\begin{figure}[t]
	\centering
	\includegraphics[width=.8\columnwidth, trim=0 520 0 0, clip]{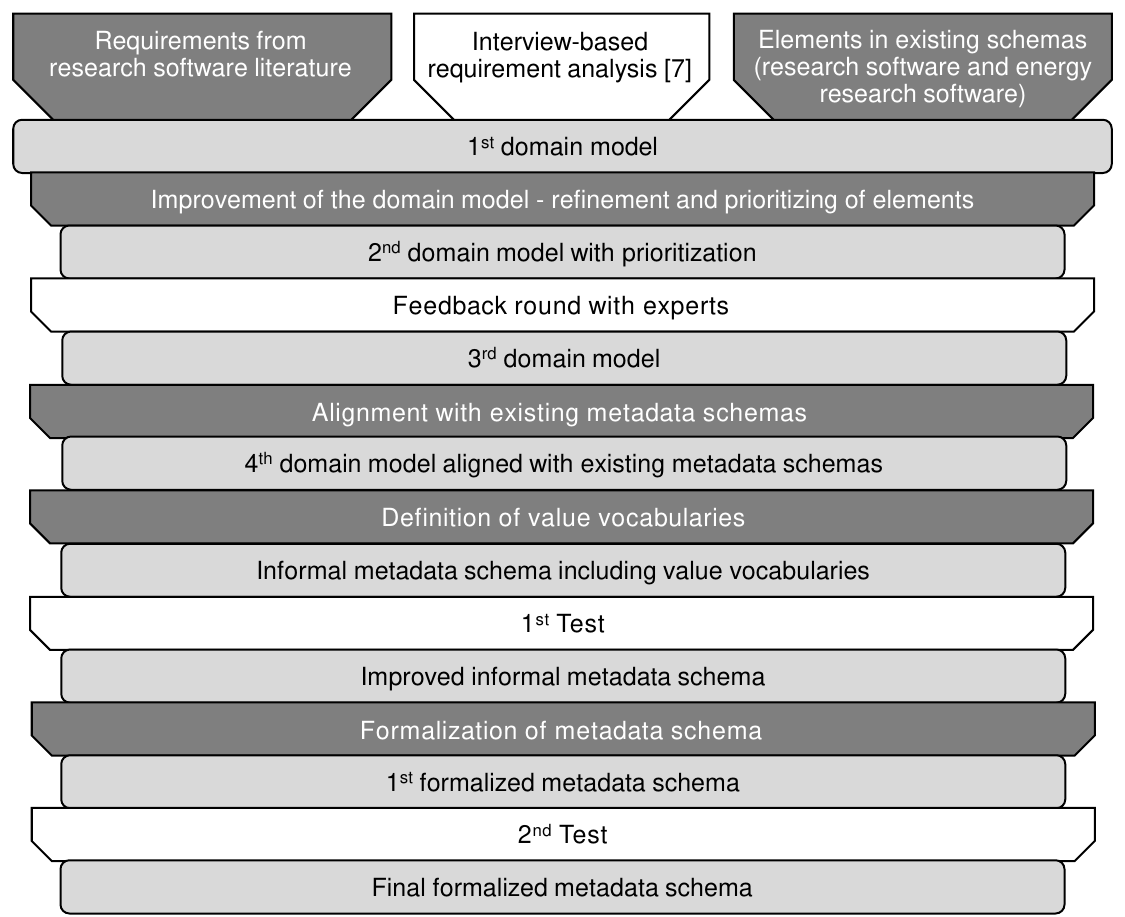}
	\caption{Overview of our method for developing the metadata schema. White boxes contain expert interaction and testing. Dark gray boxes contain development steps. Light gray boxes contain (interim) artifacts.}
    \label{fig:method}
\end{figure}

\autoref{fig:method} shows the process that we used for the development of \textit{ERSmeta}. For our method, we adapted the Me4MAP (Method for the Development of Metadata Application Profiles) proposed by  \textcite{curadomalta_Development_2017}. 
We started with gathering information from three sources: an interview-based requirements analysis, an analysis of requirements from research software literature, and an analysis of elements in existing metadata schemas. Together the requirements led to a broad domain model. In the next step, we refined and prioritized the elements in the domain model. We let multiple experts review this version of \textit{ERSmeta} to gather first feedback. Afterwards, we aligned the elements in the domain model with existing metadata schemas to achieve high interoperability. Then, we defined value vocabularies for multiple elements. We tested the resulting version of \textit{ERSmeta} with two software. Finally, we formalized the schema and performed a last test. In the following, we will provide more details for all steps.

\renewcommand{\arraystretch}{1.4}
\begin{table}[t]
    \centering
    \caption{Overview of the analyzed metadata schemas}
    \begin{tabular}{C{2.1cm} | C{8.5cm} C{2.7cm} }
         &  Software & Data \\
         \hline 
         \hline 
         General & CodeMeta \cite{jones_CodeMeta_2023}, DataDesc \cite{kuckertz_DataDesc_2024}, \href{https://cran.r-project.org/doc/
manuals/R-exts.html#The-DESCRIPTION-file}{R Package Metadata}, \href{https://packaging.python.org/en/latest/specifications/core-metadata/}{Python Package Metadata}, \gls{CFF} \cite{druskat_Citation_2021}, \href{https://research-software-directory.org/documentation/users/adding-software/}{RSD Data Model}, Software Description Ontology \cite{garijo_OKGSoft_2019} & \\
         Other domains & \textit{ontosoft} \cite{gil_OntoSoft_2016}, \textit{BiotoolsSchema} \cite{ison_BiotoolsSchema_2021} & \\
         Engineering & & m4i \cite{iglezakis_Modelling_2023} \\
         Energy & \href{https://wiki.openmod-initiative.org/}{openmod}, \href{https://openenergyplatform.org/}{\gls{OEP}~Framework and Model~Factsheets} & OEMetadata \cite{hulk_Open_2025}
    \end{tabular}
    \label{tab:analyzed_schemas}
\end{table}

\paragraph{Interview-Based Requirement Analysis} Since \textit{ERSmeta} should cover relevant domain-specific information we decided to broadly gather information requirements. Therefore, we conducted 32 expert interviews with energy researchers \cite{ferenz_Requirements_2025}. We publish our requirement analysis and the resulting domain model in \cite{ferenz_Requirements_2025}.

\paragraph{Requirements from Research Software Literature} The envisioned metadata schema should not only fulfill the current requirements by energy researchers but also should contribute to \gls{FAIR} research software. Therefore, we also analyzed the \gls{FAIR}4RS principles \cite{chuehong_FAIR_2022} and additional research software literature to identify additional requirements for \textit{ERSmeta}. Especially, \cite{gruenpeter_D44_2023} provided an important overview on these requirements. The requirements let to specific elements which we added to our domain model.

\paragraph{Elements in Existing Schemas} Additionally, we analyzed elements of related metadata schemas to identify relevant metadata elements to close gaps and specify the elements extracted from the requirements analysis. \autoref{tab:analyzed_schemas} provides an overview on the metadata schemas which we analyzed in this step. As a result, we added additional elements to our domain model.

\paragraph{Improvement of the Domain Model} We consolidated our domain model by reducing redundancies within the domain model. Also, we performed a first round of prioritization by sorting the elements into mandatory, recommended, and optional. For all elements we added definitions. If the elements were taken from existing schemas, we reused the existing definition. 

\paragraph{Feedback Round with Experts} To further improve \textit{ERSmeta}, we gathered expert feedback from five experts who work in the field of research software metadata and energy research software. The experts received the domain model with all elements, their descriptions, and the prioritization as an Excel sheet and filled a small survey to provide feedback. The goal was to identify missing elements, bad descriptions, and inconsistencies. We incorporated the received feedback in the domain model.

\paragraph{Alignment with Existing Metadata Schemas}
To achieve high interoperability, we aligned the elements in the domain model with existing metadata schemas which is called Vocabulary Alignment by \textcite{curadomalta_Development_2017}. We mapped the domain model to existing schemas including \textit{CodeMeta} \cite{jones_CodeMeta_2023}, \textit{\gls{CFF}} \cite{druskat_Citation_2021}, \textit{DataDesc} \cite{kuckertz_DataDesc_2024}, \textit{OEMetadata} \cite{hulk_Open_2025}, \textit{\href{https://openenergyplatform.org/}{\gls{OEP} Framework and Model Factsheets}}, \textit{m4i}~\cite{iglezakis_Modelling_2023}, \textit{ontosoft} \cite{gil_OntoSoft_2016}, \textit{BiotoolsSchema} \cite{ison_BiotoolsSchema_2021}, \textit{\href{http://bioschemas.org/profiles/ComputationalTool/1.0-RELEASE}{Bioschemas ComputationTool}}, and \textit{Software Description Ontology}~\cite{garijo_OKGSoft_2019}. We checked the definition of each aligned element to identify if an element from the existing ontologies or metadata schemas could be reused for \textit{ERSmeta}. We prioritized the reuse of elements from \textit{schema.org} since this is the most common ontology. If we identified a matching element we reused the name and description. Also, we added the link to the namespace. Generally, we tried to also reuse the type if it is a common data type. In some cases, we needed to change the hierarchy of our elements to better align with the existing definitions. 

\paragraph{Definition of Value Vocabularies}
To limit the allowed values for certain elements and, therefore, increase the interoperability of the produced metadata, we defined value vocabularies for multiple elements. We used different sources for the value vocabularies. If possible, we used existing ontologies like the \textit{Open Energy Ontology} \cite{booshehri_Introducing_2021} or semantic knowledge bases like \textit{wikidata}. Otherwise, we  created a list of allowed terms from literature or other existing sources. For some elements, we extended the existing lists and for other elements, we constructed complete lists from our expert knowledge.

\paragraph{1\textsuperscript{st} Test with two Software} To further improve \textit{ERSmeta}, we tested its usability by creating metadata based on the metadata schema for two energy research software. For each software, the metadata were created together with a developer of the software. Based on this testing, the schema was improved (e.g., descriptions, allowed values, etc.).

\paragraph{Formalization of Metadata Schema}
To formalize \textit{ERSmeta}, we decided to use two syntax types. We used the \href{https://www.w3.org/TR/shacl/}{\gls{SHACL}} \cite{knublauch_Shapes_2017} in turtle to allow high semantic web compatibility. \gls{SHACL} allows complex constrains for all elements, e.g., an element can be limited to a certain class which is useful for dynamic value vocabularies. Additional, we used JSON-LD Schema to allow easier human readability as well as better interoperability to \textit{CodeMeta} which is also provided as JSON-LD Schema. In this formalization only the element vocabulary is semantic web compatible but not the constrains or values. We started with defining the \gls{SHACL}. Afterwards, we translated it to JSON-LD Schema and added additional required information to the JSON-LD Schema. Also, we added crosswalks to other metadata schemas into our repository.

\paragraph{2\textsuperscript{nd} Test with two Software}
We performed a second test with the same two software to also test the formalization. Based on the informal metadata from the first test round, formal metadata were created as turtle following the \gls{SHACL} and as JSON-LD following the JSON-LD Schema. We tested if all information could be expressed in both formats. Also, we evaluated if the validation against the schema works successfully. We were able to fix some errors and improve the descriptions.

\subsection*{Evaluation Design}
\label{subsec:eval_method}
After the development of \textit{ERSmeta}, we aimed for a broader evaluation. Since our development already incorporated requirements from the literature on \gls{FAIR} research software (e.g., contributor role) and the needs of researchers looking for energy research software, we concentrated on the aspect of creating metadata based on the metadata schema. Metadata for energy research software should generally be created by the developers of the respective software, e.g., research software engineers or researchers who code. To address our evaluation aspect, we designed our evaluation study as in-vivo experiment, where participants create metadata for their own energy research software and share their experience in a structured way via a questionnaire. Questionnaires are an often used approach to evaluate metadata schemas \cite{palavitsinis_Evaluation_2009}.
The evaluation study was approved by the OFFIS study board (approval number: 2025E021).

For the creation of the metadata, we used our tool \gls{SMECS} \cite{ferenz_SMECS_2025b, ferenz_Software_2025}. \gls{SMECS} is a web-tool with a light backend that is able to extract structured metadata from GitHub and GitLab repositories, such as name and keywords, based on a provided URL. \gls{SMECS} presents the extracted metadata and other metadata elements in an user-friendly interface where the user can curate the extracted metadata and enter additional metadata. The created metadata can be downloaded as JSON file. In this way, \gls{SMECS} provides semi-automated creation of metadata. The first version of \gls{SMECS} is based on \textit{CodeMeta}. We integrated \textit{ERSmeta}~(V0.8) into \gls{SMECS} to allow users to create domain-specific metadata for their energy research software.

\begin{figure}[]
	\centering
	\includegraphics[width=.8\columnwidth, trim=0 750 0 10, clip]{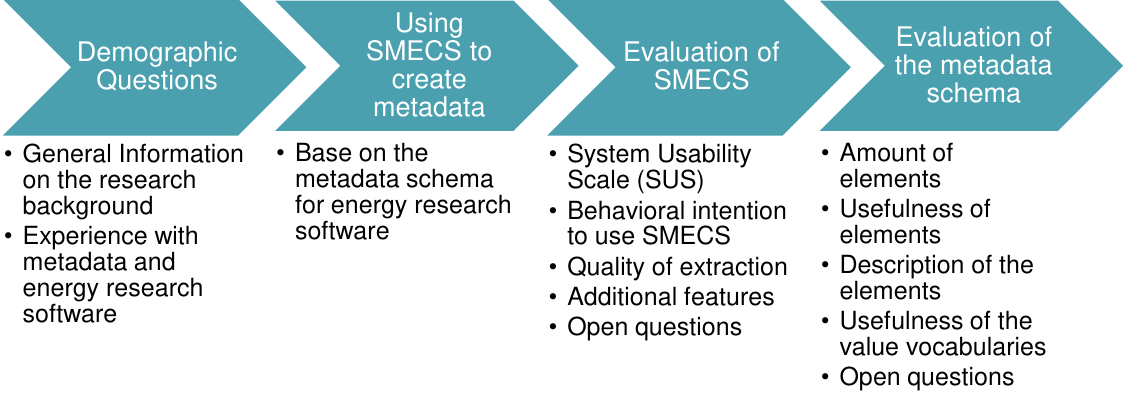}
	\caption{Overview of our questionnaire for the evaluation of the metadata schema}
    \label{fig:method_eval}
\end{figure}

We designed the questionnaire to gather the participants' experience with creating metadata using \gls{SMECS} and \textit{ERSmeta}. The questionnaire is available on Zenodo \cite{ferenz_Questionnaire_2025}. Its overall structure is presented in \autoref{fig:method_eval} and consists of several sections.

First, the participants were asked to provide general demographic information, such as how long they have been in research, as well as information on their experience with metadata and energy research software. This is followed by a section where the participants were asked to open \gls{SMECS} and create metadata for their own energy research software. Afterwards, they were asked to answer questions on their experience with \gls{SMECS}, including a \gls{SUS} \cite{brooke_sus_1995}, questions on the behavioral intention to use \gls{SMECS} based on the \gls{UTAUT}~\cite{venkatesh_User_2003} and \gls{UTAUT}3~\cite{farooq_Acceptance_2017}, and on the quality of the extraction based on \gls{TAM}2~\cite{venkatesh_Theoretical_2000}. Also, we asked the participants on their opinion for certain features ideas and added open questions to allow free text answers on the usage of \gls{SMECS} including the option to provide suggestions for further improvements.

Afterwards, the participants were asked to evaluate their experience with \textit{ERSmeta}, including the amount and distribution of elements, the usefulness of elements, the descriptions of the elements, and the usefulness of the value vocabularies. For the different categories, we created question items based on \gls{TAM} \cite{davis_Perceived_1989}, \gls{TAM}2 \cite{venkatesh_Theoretical_2000}, \gls{TAM}3 \cite{venkatesh_Technology_2008}, \gls{UTAUT} \cite{venkatesh_User_2003}, and \gls{UTAUT}2 \cite{venkatesh_User_2003}. We also added some open questions to allow the participants to add free text, e.g., to propose additional elements. Finally, we also asked the participants to provide us the created metadata for further evaluation. We used a separate form for this purpose to avoid linking between the answers to the questionnaire and the provided metadata.

We invited participants via different mailing lists (\href{https://www.strommarkttreffen.org/english/}{Strommarkttreffen}, Newsletter \href{https://nfdi4energy.uol.de/}{NFDI4Energy}, \href{https://groups.google.com/g/openmod-initiative}{openmod}, social media (\href{https://www.linkedin.com/company/nfdi4energy/}{NFDI4Energy on LinkedIn}) and directly contacted developers of known energy research software to motivate them to participate in the study.

%% file: content/results.tex
\begin{figure}[]
	\centering
	\includegraphics[width=\columnwidth, trim=0 300 0 0, clip]{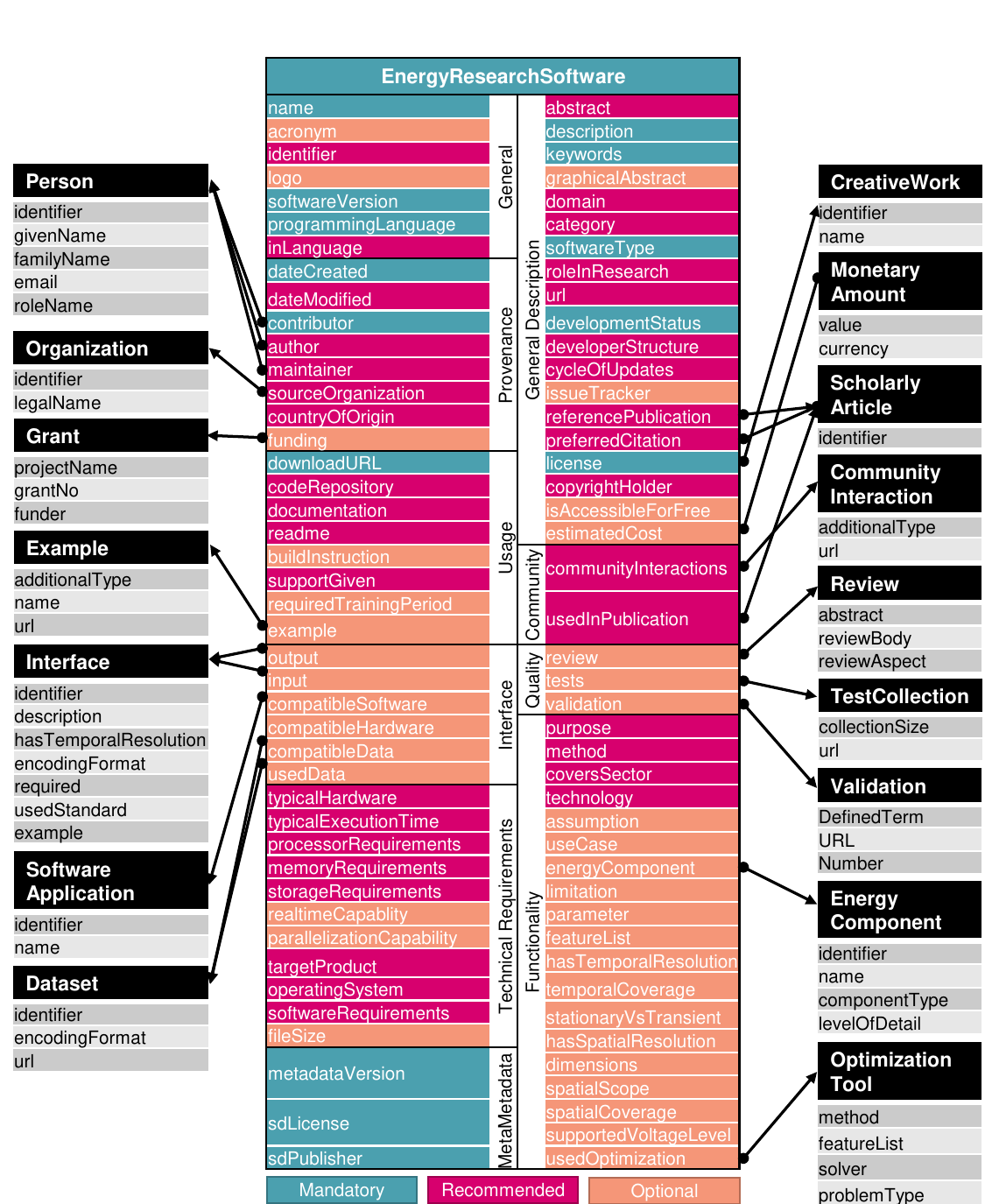}
	\caption{\textit{ERSmeta}: metadata schema for energy research software (before the evaluation)}
    \label{fig:schema}
\end{figure}

\section*{Metadata Schema: ERSmeta}
\label{subsec:schema}

The developed metadata schema for energy research software (\textit{ERSmeta}) is shown in \autoref{fig:schema}. The schema is available as a JSON Schema in JSON-LD and as \gls{SHACL} in turtle on GitHub \cite{ferenz_ERSmeta_2025} and Zenodo \cite{ferenz_ERSmeta_2025a} allowing for simple reuse.

The schema consists of 86 top level elements clustered into ten thematic areas which help to get a better overview of the schema. The thematic areas cover different aspects of a software, e.g., the area functionality focuses on domain-specific aspects of the functionality of a software.
23 elements of the top level elements use value types that require schemas by themselves, e.g., \texttt{author}. For this purpose, additional 16 metadata schemas with 48 elements are defined, shown left and right in \autoref{fig:schema}. When taking into account that each of the top level elements needs to be filled at least once, overall 158 elements are presented in the metadata schema for energy research software.

\begin{figure}[]
	\centering
	\includegraphics[width=.7\columnwidth, trim=60 610 60 40, clip]{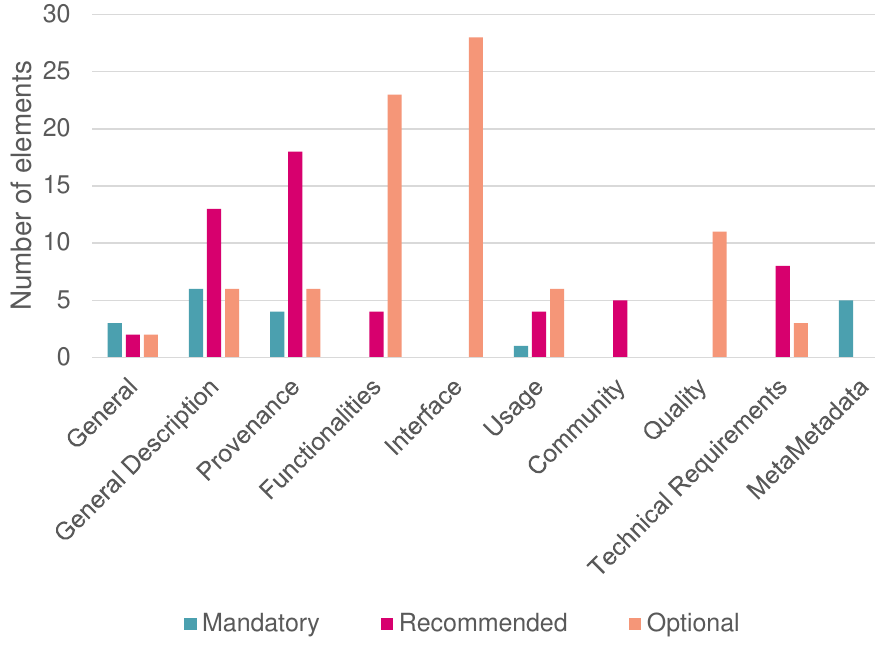}
	\caption{Distribution of elements between mandatory, required, and optional}
    \label{fig:schema_mandatory}
\end{figure}
To provide orientation on the importance of the elements, we sorted them into three categories: mandatory, recommended, and optional. The 19 mandatory elements are required to be filled. They are a good way to ensure a certain quality level of the metadata. Examples include the \texttt{name} and the \texttt{programmingLanguage} of a software. The 54 recommended elements provide a good overview of the software and, therefore, should be filled by the users of the schema. \texttt{referencePublication} and \texttt{purpose} are examples for recommended elements. The 85 optional elements are useful but less important to describe the software, such as \texttt{funding}, \texttt{input}, and \texttt{output}. \autoref{fig:schema_mandatory} provides an overview of how the mandatory, recommended, and optional elements are distributed over the ten thematic areas.

\begin{figure}[t]
\centering
\begin{subfigure}[b]{0.48\textwidth}
 	\includegraphics[width=1\textwidth, trim=50 755 120 25, clip]{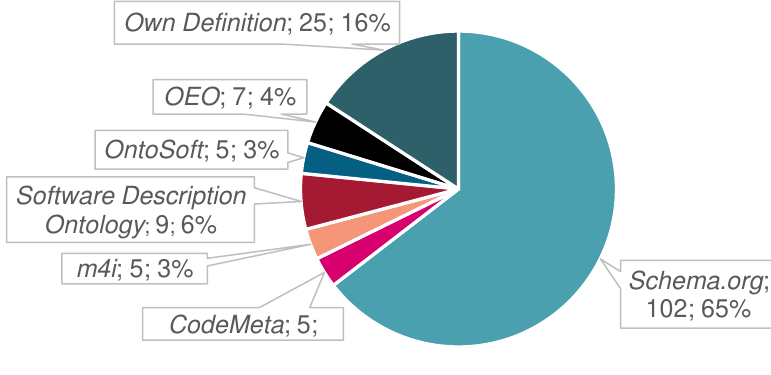}
 	\caption{Reuse of elements}
     \label{fig:schema_reuse}
\end{subfigure}
\hfill
\begin{subfigure}[b]{0.48\textwidth}
	\includegraphics[width=1\textwidth, trim=120 750 80 30, clip]{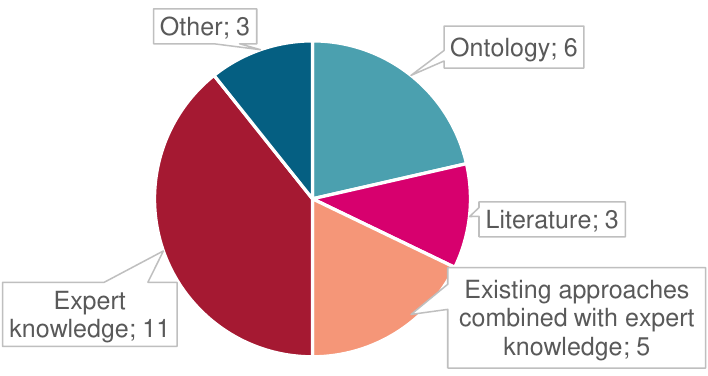}
	\caption{Sources for value vocabularies}
    \label{fig:schema_value_voc}
\end{subfigure}
\caption{Statistics of reused elements and used value vocabularies in \textit{ERSmeta}}
\label{fig:figure}
\end{figure}

We reused many existing elements to achieve high interoperability to other metadata schemas. \autoref{fig:schema_reuse} gives an overview of how many elements were reused from which existing metadata schema or ontology. Most elements were reused from \textit{schema.org}, which is a highly used ontology primarily introduced to describe resources in the web for search engines. As introduced in the related work, \textit{CodeMeta} \cite{jones_CodeMeta_2023} is the most common metadata schema for research software and is based on \textit{schema.org}. Reused elements which are part of \textit{CodeMeta} and \textit{schema.org} are counted as \textit{schema.org}. We also reused some elements from \textit{CodeMeta} that are not yet part of \textit{schema.org}, such as \texttt{developmentStatus}. Additional elements on general aspects of research software were added from the \textit{Software Description Ontology} \cite{garijo_Software_2021}, e.g., \texttt{preferredCitation}, and \textit{OntoSoft} \cite{gil_OntoSoft_2016}, e.g., \texttt{compatibleSoftware}. More domain-specific aspects were added from the \textit{Open Energy Ontology} \cite{booshehri_Introducing_2021}, such as \texttt{coversSector}, and from \textit{m4i} \cite{iglezakis_Modelling_2023}, such as \texttt{method}. Additionally, 25 new elements were defined covering general aspects like \texttt{developerStructure} or domain-specific aspects like \texttt{energyComponent}.

To increase the interoperability of the created metadata, we integrated value vocabularies for multiple elements. Value vocabularies limit the allowed values for a certain element to a controlled vocabulary, which can be a predefined list, an ontology, or a specific class of an ontology \cite{zeng_Metadata_2022}. They enable to create metadata which are easy to compare and filter. We used different sources for the value vocabularies which are displayed in \autoref{fig:schema_value_voc}. We included existing concepts from ontologies like the \textit{Open Energy Ontology} \cite{booshehri_Introducing_2021} and semantic name authorities like \textit{wikidata}. We also used categorizations from literature, e.g., for \texttt{roleInReseearch} from \textcite{hasselbring_MultiDimensional_2025} and for \texttt{realtimeCapability} from \textcite{shin_Realtime_1994}. For some elements, we also extended the existing sources with our own expert knowledge, e.g., for \texttt{softwareType} we used and extended a value vocabulary from \textit{BiotoolsSchema} \cite{ison_BiotoolsSchema_2021}. Additionally, we created some predefined lists of terms with our own expert knowledge, e.g., \texttt{supportedVoltageLevel}. 

\section*{Evaluation of the Metadata Schema}
\label{subsec:eval}

We evaluated \textit{ERSmeta} through a survey as introduced in the method section. 
The survey answers are published in \cite{ferenz_Dataset_2026} while the created metadata of the survey participants are not openly available due to privacy constrains. The analysis scripts, which were also used to create the Figures in this section, are published in \cite{ferenz_ERSmeta_2025b}.

\begin{figure}[t]
	\centering
	\includegraphics[width=\columnwidth, trim=0 0 0 0, clip]{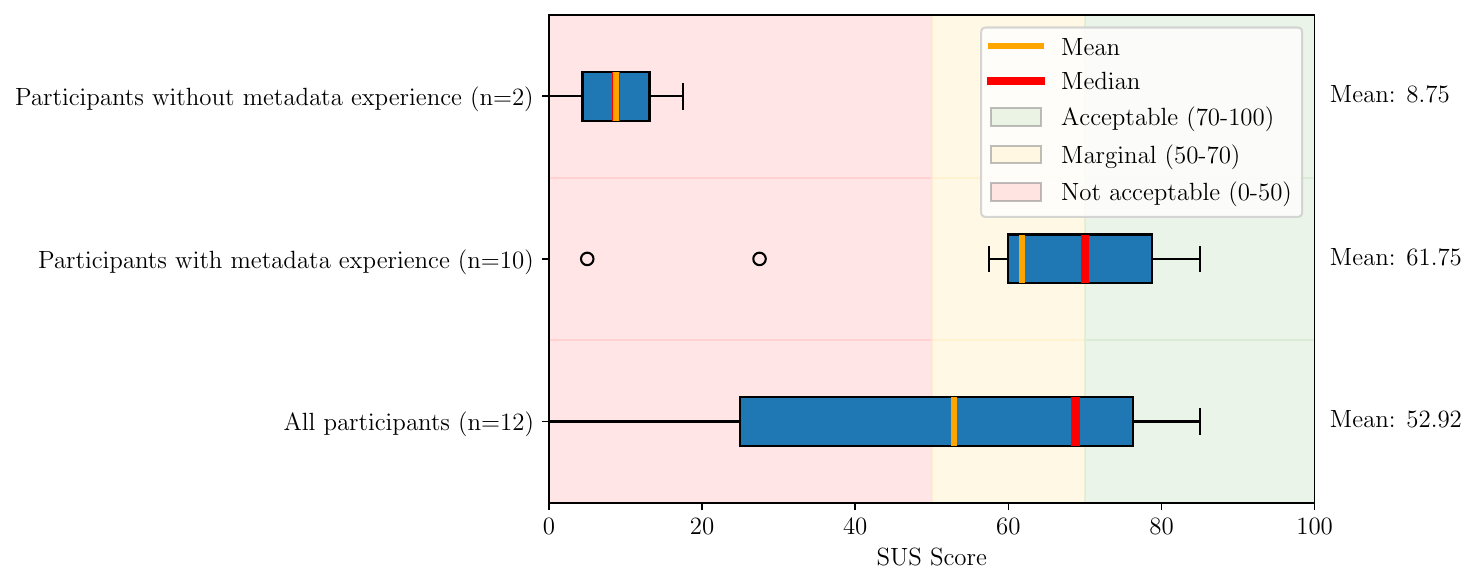}
	\caption{\acrfull{SUS} scores for participants with and without familiarity of metadata schemas and total. }
    \label{fig:sus}
\end{figure}

The survey participants consisted of twelve individuals (n=12) who completed the entire questionnaire, with ten of them providing the created metadata for further analysis. The participants held diverse positions within their research institutions, including research-supporting staff, PhD students, postdoctoral researchers, and other research staff. Their backgrounds spanned various domains, with one participant from economics, five from computer science, and six from engineering.
The participants exhibited varying levels of experience with metadata schemas, with two individuals having no prior familiarity with metadata schemas. Notably, these two participants were more critical of the tooling and metadata schema. For instance, the \gls{SUS} scores, presented in \autoref{fig:sus}, demonstrate a difference between participants with and without metadata experience. This trend is also observed in other survey items, suggesting that a lack of understanding or motivation for the metadata schema may contribute to a more negative perception of \textit{ERSmeta}.

\begin{figure}[t]
	\centering
	\includegraphics[width=\columnwidth, trim=0 0 0 0, clip]{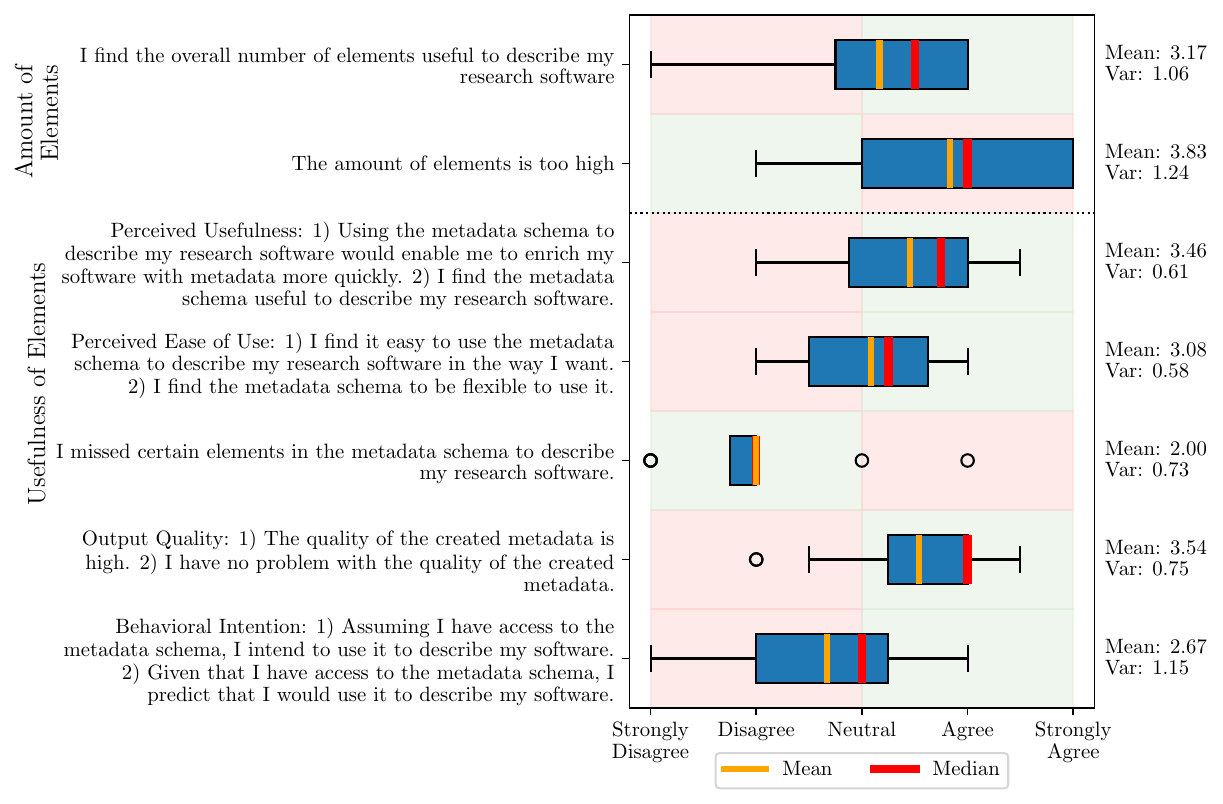}
	\caption{Impressions on the amount of elements and the usefulness of the elements (n=12). For the usefulness of elements, items from the construct (questions aiming for the same aspect) are averaged. For means and variances, the likert scale is mapped to numbers (strongly disagree = 1, ... , strongly agree = 5) }
    \label{fig:elements}
\end{figure}

The \gls{SUS} scores presented in \autoref{fig:sus} suggest that the participants found the system in average usable but its usability can be further improved. Specifically, the survey questions (items) related to complexity ("I found the system unnecessarily complex") and the desire for frequent use ("I think that I would like to use this system frequently") contributed significantly to the lower \gls{SUS} scores.
We attribute this criticism primarily to the high number of elements in \textit{ERSmeta}, as the open text fields revealed that participants were concerned with the high number of elements and the time-consuming nature of the metadata creation process. This is also reflected in the participants' responses to the items related to the amount of elements, as shown in \autoref{fig:elements}. Notably, the participants agreed that "The amount of elements is too high", indicating that users need less elements to be filled. Meanwhile the participants also partly agreed that the overall number of elements is useful to describe their research software.

\begin{figure}[p]
	\centering
	\includegraphics[width=\columnwidth, trim=0 0 0 0, clip]{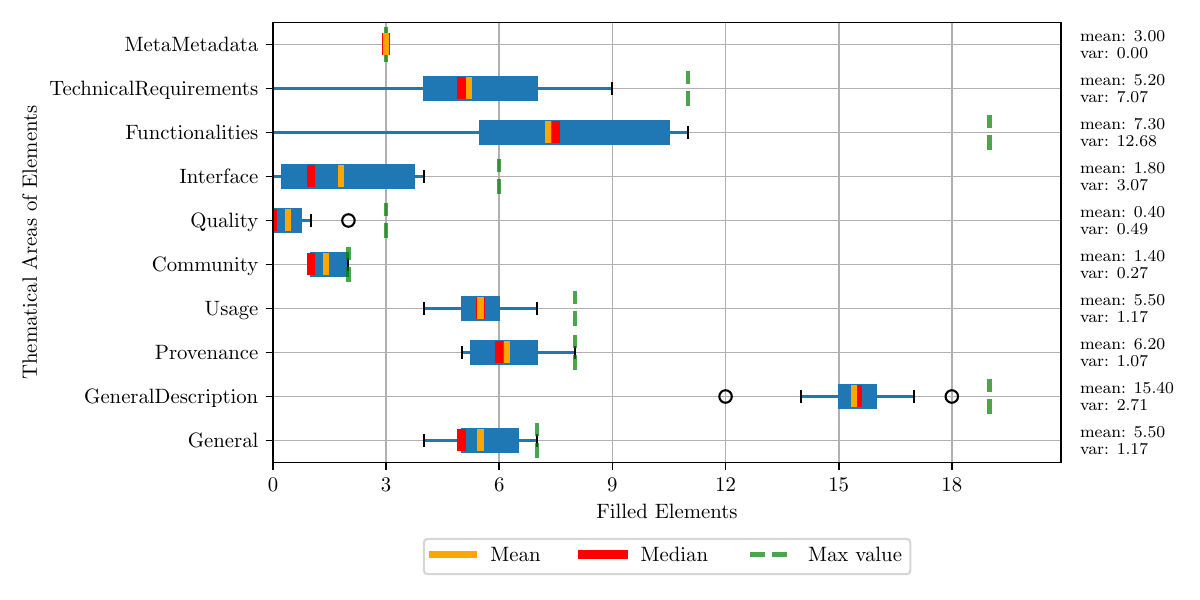}
	\caption{Filled elements per thematical area of the ten metadata sets}
    \label{fig:themetical_area}
\end{figure}

However, despite the criticisms of the amount of elements in the metadata schema, the participants found the elements to be useful and relevant for describing their energy research software. As shown in \autoref{fig:elements}, the participants rated the usefulness of the elements as positive, indicating that they were effective in capturing the necessary information.
The participants also ranked the quality of the created metadata as high. This is further supported by the open text fields, where participants mentioned that the metadata were relevant and helpful.
When analyzing the submitted metadata sets, it was found that only three elements (\texttt{typicalExecutionTime}, \texttt{example}, and \texttt{estimatedCosts}) were not used for any software. This supports that the participants found the majority of the elements to be relevant and useful.
When asked if elements were still missing, the participants mainly disagreed, indicating that the current set of elements was sufficient for their needs. When asked to specify missing elements in the open text fields, the participants only mentioned four properties (related publication, related projects, origin of data, and intended public). Of these, two are already included in the schema, one is out-of-scope (metadata for data), and only the fourth may be added to the metadata schema.
The distribution of elements across different thematic areas is shown in \autoref{fig:themetical_area}. Notably, elements related to functionalities were often not filled, which we assume is due to the fact that some elements may not be applicable to every energy research software.

\begin{figure}[p]
	\centering
	\includegraphics[width=\columnwidth, trim=0 0 0 0, clip]{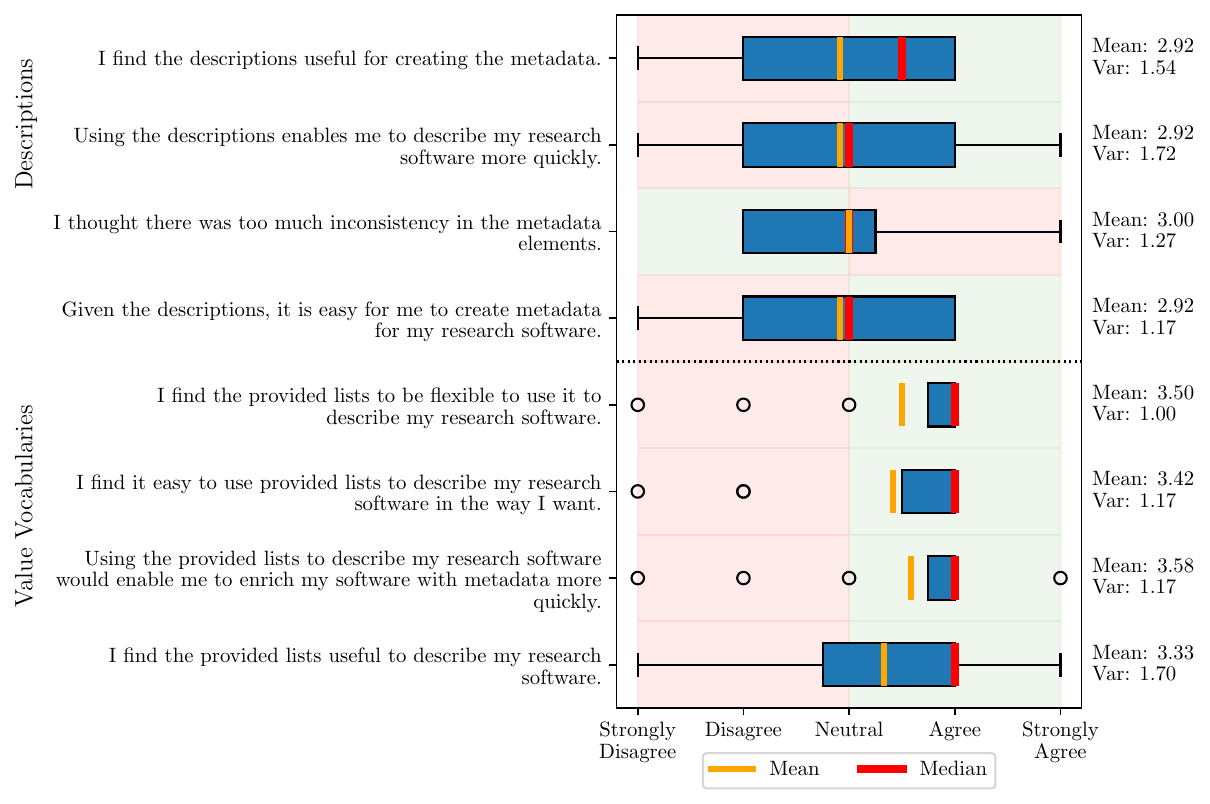}
	\caption{Usefulness of the descriptions and value vocabularies (n=12)}
    \label{fig:description_value_voc}
\end{figure}

The usefulness of the descriptions of the elements is presented in \autoref{fig:description_value_voc}. The participants' ratings indicate a neutral average, suggesting that the descriptions are a good starting point but require further improvement. Specifically, the participants noted in the open text fields that the descriptions would benefit from more examples to enhance the usability of \textit{ERSmeta}.

\autoref{fig:description_value_voc} also presents the usefulness of the value vocabularies, which were found to be effective and helpful by the participants. The participants rated the usefulness of the value vocabularies as positive, indicating that they were a valuable asset to \textit{ERSmeta}. Furthermore, the value vocabularies were highlighted as a positive aspect of the metadata schema in the open text fields. While the participants were generally satisfied with the value vocabularies, they did suggest that minor extensions would be necessary to make them more comprehensive.

The evaluation of \textit{ERSmeta} reveals a mixed assessment from the participants. While they found the elements to be useful and relevant, the high number of elements and complexity of the system were significant drawbacks. Nevertheless, the results of this evaluation will serve as a valuable foundation for refining and improving the metadata schema and the software, ultimately enhancing its ability to fully meet the needs of researchers and research software engineers in the energy research software community.

%% file: content/discussion.tex
\section*{Discussion}
\label{sec:dis}

The development and evaluation of \textit{ERSmeta} has provided valuable insights into the challenges and opportunities of creating a metadata schema for this domain. However, the study also has some limitations that should be acknowledged.

One of the limitations is the limited geographic scope of the evaluation, with most participants coming from Germany. Also the coverage of different energy research software with respect to research topic and size is limited in the evaluation, e.g., with most participants being part of the development of bigger energy research software. Since energy research is a highly interdisciplinary field it was a challenge to include all required perspectives. While we have already included diverse perspectives in the requirement analysis in \cite{ferenz_Requirements_2025}, it may not have been enough to cover all parts of energy research with sufficient depth.

Additionally, the evaluation of the tooling (\gls{SMECS}) and the metadata schema simultaneously may have introduced a confounding factor, making it challenging to isolate the participants' experiences. Although we attempted to mitigate this by first asking about \gls{SMECS}, where their experience was more direct, the impressions of \gls{SMECS} and \textit{ERSmeta} became intertwined. Specifically, the participants' perception of \textit{ERSmeta} was heavily influenced by its representation in \gls{SMECS}, which in turn was shaped by \textit{ERSmeta}.

Despite the limitations of the evaluation, \textit{ERSmeta} has demonstrated potential in supporting the creation and management of metadata for energy research software. Notably, the survey participants did not identify any significant gaps in the schema, with nearly no additional elements being suggested. Furthermore, the descriptions of the elements were found to be usable, and the value vocabularies were particularly well-received by the survey participants, who highlighted them as a positive aspect of \textit{ERSmeta}.

The representation of the metadata schema has shown to be a critical factor in its usability for energy researchers. While the current tool displays all elements to the user, a reduction of the shown elements to the most essential and best fitting aspects is a crucial next step. Additionally, it would be beneficial to explore the possibility of varying the number of shown elements based on the amount of extracted information, potentially leading to a more streamlined and user-friendly experience.

\textit{ERSmeta} mainly defines new domain-specific elements, but also some non-domain specific elements, such as \texttt{communityInteraction}. It is worth to carefully analyze whether these non-domain-specific elements are relevant to generally describe research software and, if so, to add them to general standards like \textit{CodeMeta}. 

\textit{ERSmeta} is a formalized schema which makes it more flexible than the existing not formalized approaches in the energy domain which we presented in the related work. Also, we showed in our evaluation
that \textit{ERSmeta} reuses many existing elements from established metadata schemas and ontologies which makes the created metadata interoperable with these approaches. \textit{ERSmeta} is based on a clear process which we outlined in the section on the development of the schema.
However, further evaluation is required to deeper assess the usability and effectiveness of the schema. One potential area of future research is the integration of the metadata schema in a software registry, ideally following best practices as described in \cite{garijo_Nine_2022}. This would allow a more detailed evaluation of the usability of the schema and its ability to support the search and discovery of energy research software.

Future developments in general metadata for research software need to be carefully observed to see if further changes in \textit{ERSmeta} are necessary. For example, \textit{ERSmeta} does not describe interfaces in a good machine-readable way, which is a challenge if \textit{ERSmeta} should be human usable (and creatable) at the same time. \textcite{kuckertz_DataDesc_2024} described an approach, where the interface information is included in the source code and then extracted automatically into the metadata. This could be a potential solution to this challenge. Since \textit{ERSmeta} is largely built on \textit{CodeMeta}, it should be possible to quickly adapt new approaches to \textit{ERSmeta}.

In conclusion, \textit{ERSmeta} is an important first step towards creating a metadata standard for energy research software. For the development of a standard, additional research is needed to evaluate the usability of the metadata schema, especially within a software registry. However, \textit{ERSmeta} is ready to be used to describe energy research software and, therefore, contributes to one important aspect to enable  \gls{FAIR} energy research software. 

%% file: Literature.bib
@article{barker_Introducing_2022,
  title = {Introducing the {{FAIR Principles}} for Research Software},
  author = {Barker, Michelle and Chue Hong, Neil P. and Katz, Daniel S. and Lamprecht, Anna-Lena and {Martinez-Ortiz}, Carlos and Psomopoulos, Fotis and Harrow, Jennifer and Castro, Leyla Jael and Gruenpeter, Morane and Martinez, Paula Andrea and Honeyman, Tom},
  year = {2022},
  month = oct,
  journal = {Scientific Data},
  volume = {9},
  number = {1},
  pages = {622},
  issn = {2052-4463},
  doi = {10.1038/s41597-022-01710-x},
  copyright = {2022 The Author(s)},
  langid = {english}
}

@article{booshehri_Introducing_2021,
  title = {Introducing the {{Open Energy Ontology}}: {{Enhancing}} Data Interpretation and Interfacing in Energy Systems Analysis},
  shorttitle = {Introducing the {{Open Energy Ontology}}},
  author = {Booshehri, Meisam and Emele, Lukas and Fl{\"u}gel, Simon and F{\"o}rster, Hannah and Frey, Johannes and Frey, Ulrich and Glauer, Martin and Hastings, Janna and Hofmann, Christian and {Hoyer-Klick}, Carsten and H{\"u}lk, Ludwig and Kleinau, Anna and Knosala, Kevin and Kotzur, Leander and Kuckertz, Patrick and Mossakowski, Till and Muschner, Christoph and Neuhaus, Fabian and Pehl, Michaja and Robinius, Martin and Sehn, Vera and Stappel, Mirjam},
  year = {2021},
  month = sep,
  journal = {Energy and AI},
  volume = {5},
  pages = {100074},
  issn = {2666-5468},
  doi = {10.1016/j.egyai.2021.100074},
  langid = {english}
}

@incollection{brooke_sus_1995,
  title = {{{SUS}}: {{A}} Quick and Dirty Usability Scale},
  shorttitle = {{{SUS}}},
  booktitle = {Usability {{Eval}}. {{Ind}}.},
  author = {Brooke, John},
  year = {1995},
  month = nov,
  edition = {1st Edition},
  volume = {189},
  publisher = {CRC Press},
  doi = {10.1201/9781498710411-35},
  isbn = {978-0-429-15701-1}
}

@techreport{chuehong_FAIR_2022,
  title = {{{FAIR Principles}} for {{Research Software}} ({{FAIR4RS Principles}})},
  author = {Chue Hong, Neil P and Katz, Daniel S and Barker, Michelle and Lamprecht, Anna-Lena and Martinez, Carlos and Psomopoulos, Fotis E and Harrow, Jen and Castro, Leyla Jael and Gruenpeter, Morane and Martinez, Paula Andrea and Honeyman, Tom and Struck, Alexander and Lee, Allen and Loewe, Axel and {van Werkhoven}, Ben and Jones, Catherine and Garijo, Daniel and Plomp, Esther and Genova, Francoise and Shanahan, Hugh and Leng, Joanna and Hellstr{\"o}m, Maggie and Sinha, Manodeep and Kuzak, Mateusz and Herterich, Patricia and Zhang, Qian and Islam, Sharif and Sansone, Susanna-Assunta and Pollard, Tom and Atmojo, Udayanto Dwi and Williams, Alan and Czerniak, Andreas and Niehues, Anna and Fouilloux, Anne Claire and Desinghu, Bala and Richard, C{\'e}line and Gray, Charles and Erdmann, Chris and N{\"u}st, Daniel and Tartarini, Daniele and Anzt, Hartwig and Todorov, Ilian and McNally, James and Moldon, Javier and Burnett, Jessica and Belhajjame, Khalid and Sesink, Laurents and Hwang, Lorraine and Roberto, Marcos and Wilkinson, Mark D and Servillat, Mathieu and Liffers, Matthias and Fox, Merc and Lynch, Nick and Lavanchy, Paula Martinez and Gesing, Sandra and Stevens, Sarah and Cuesta, Martinez and Peroni, Silvio and {Soiland-Reyes}, Stian and Bakker, Tom and Rabemanantsoa, Tovo and Sochat, Vanessa and Yehudi, Yo},
  year = {2022},
  month = may,
  pages = {32},
  institution = {FAIR4RS WG},
  doi = {10.15497/RDA00068},
  langid = {english}
}

@incollection{curadomalta_Development_2017,
  title = {The {{Development}} Process of a {{Metadata Application Profile}} for the {{Social}} and {{Solidarity Economy}}},
  booktitle = {Developing {{Metadata Application Profiles}}},
  author = {Curado Malta, Mariana and Baptista, Ana Alice},
  editor = {Malta, Mariana Curado and Baptista, Ana Alice and Walk, Paul},
  year = {2017},
  pages = {98--117},
  publisher = {IGI Global},
  address = {Hershey, PA, USA},
  doi = {10.4018/978-1-5225-2221-8.ch005},
  isbn = {978-1-5225-2221-8}
}

@article{davis_Perceived_1989,
  title = {Perceived {{Usefulness}}, {{Perceived Ease}} of {{Use}}, and {{User Acceptance}} of {{Information Technology}}},
  author = {Davis, Fred D.},
  year = {1989},
  journal = {MIS Quarterly},
  volume = {13},
  number = {3},
  eprint = {249008},
  eprinttype = {jstor},
  pages = {319--340},
  issn = {0276-7783},
  doi = {10.2307/249008}
}

@techreport{druskat_Citation_2021,
  title = {Citation {{File Format}}},
  author = {Druskat, Stephan and Spaaks, Jurriaan H. and Chue Hong, Neil and Haines, Robert and Baker, James and Bliven, Spencer and Willighagen, Egon and {P{\'e}rez-Su{\'a}rez}, David and Konovalov, Alexander},
  year = {2021},
  month = aug,
  institution = {Zenodo},
  url = {https://zenodo.org/records/5171937}
}

@article{farooq_Acceptance_2017,
  title = {Acceptance and Use of Lecture Capture System ({{LCS}}) in Executive Business Studies: {{Extending UTAUT2}}},
  shorttitle = {Acceptance and Use of Lecture Capture System ({{LCS}}) in Executive Business Studies},
  author = {Farooq, Muhammad Shoaib and Salam, Maimoona and Jaafar, Norizan and Fayolle, Alain and Ayupp, Kartinah and {Radovic-Markovic}, Mirjana and Sajid, Ali},
  year = {2017},
  month = nov,
  journal = {Interactive Technology and Smart Education},
  volume = {14},
  number = {4},
  pages = {329--348},
  publisher = {Emerald Publishing Limited},
  issn = {1741-5659},
  doi = {10.1108/ITSE-06-2016-0015},
  langid = {english}
}

@misc{ferenz_ERSmeta_2025,
  title = {{{ERSmeta}}},
  author = {Ferenz, Stephan},
  year = {2025},
  month = mar,
  url = {https://github.com/NFDI4Energy/ERSmeta},
  urldate = {2025-04-15},
  copyright = {CC0-1.0}
}

@article{ferenz_Improved_2023,
  title = {Towards {{Improved Findability}} of {{Energy Research Software}} by {{Introducing}} a {{Metadata-based Registry}}},
  author = {Ferenz, Stephan and Nie{\ss}e, Astrid},
  year = {2023},
  month = nov,
  journal = {ing.grid},
  volume = {1},
  number = {2},
  publisher = {Universit{\"a}ts- und Landesbibliothek Darmstadt},
  issn = {2941-1300},
  doi = {10.48694/inggrid.3837},
  langid = {english}
}

@article{ferenz_Requirements_2025,
  title = {Requirements for a {{Metadata Scheme}} to {{Enable FAIR Energy Research Software}}},
  author = {Ferenz, Stephan and Werth, Oliver and Nie{\ss}e, Astrid},
  year = {2025},
  month = feb,
  journal = {Electronic Communications of the EASST},
  volume = {83},
  issn = {1863-2122},
  doi = {10.14279/eceasst.v83.2594},
  copyright = {Copyright (c) 2025 Stephan Ferenz, Oliver Werth, Astrid Nie{\ss}e},
  langid = {english}
}

@article{ferenz_SMECS_2025b,
  title = {{{SMECS}}: {{A Software Metadata Extraction}} and {{Curation Software}}},
  shorttitle = {{{SMECS}}},
  author = {Ferenz, Stephan and Jafarbigloo, Aida and Werth, Oliver and Nie{\ss}e, Astrid},
  year = 2025,
  month = dec,
  journal = {Electronic Communications of the EASST},
  volume = {85},
  issn = {1863-2122},
  doi = {10.14279/eceasst.v85.2708},
  urldate = {2025-12-16},
  copyright = {Copyright (c) 2025 Stephan Ferenz, Aida Jafarbigloo, Oliver Werth, Astrid Nie\ss e},
  langid = {english}
}

@misc{ferenz_Software_2025,
  title = {Software {{Metadata Extraction}} and {{Curation Software}} ({{SMECS}})},
  author = {Ferenz, Stephan and Jafarbigloo, Aida},
  year = {2025},
  month = may,
  url = {https://github.com/NFDI4Energy/SMECS},
  urldate = {2025-03-18},
  copyright = {AGPL-3.0},
  howpublished = {Carl von Ossietzky Universit{\"a}t Oldenburg}
}

@article{garijo_Nine_2022,
  title = {Nine Best Practices for Research Software Registries and Repositories},
  author = {Garijo, Daniel and M{\'e}nager, Herv{\'e} and Hwang, Lorraine and Trisovic, Ana and Hucka, Michael and Morrell, Thomas and Allen, Alice},
  year = {2022},
  month = aug,
  journal = {PeerJ Computer Science},
  volume = {8},
  pages = {e1023},
  issn = {2376-5992},
  doi = {10.7717/peerj-cs.1023},
  langid = {english}
}

@inproceedings{garijo_OKGSoft_2019,
  title = {{{OKG-Soft}}: {{An Open Knowledge Graph}} with {{Machine Readable Scientific Software Metadata}}},
  shorttitle = {{{OKG-Soft}}},
  booktitle = {2019 15th {{International Conference}} on {{eScience}} ({{eScience}})},
  author = {Garijo, Daniel and Osorio, Maximiliano and Khider, Deborah and Ratnakar, Varun and Gil, Yolanda},
  year = {2019},
  month = sep,
  pages = {349--358},
  doi = {10.1109/eScience.2019.00046}
}

@misc{garijo_Software_2021,
  title = {The {{Software Description Ontology}}},
  author = {Garijo, Daniel and Ratnakar, Varun and Gil, Yolanda and Khider, Deborah},
  year = {2021},
  month = may,
  url = {https://w3id.org/okn/o/sd/1.9.0},
  urldate = {2025-07-25}
}

@inproceedings{gil_OntoSoft_2015,
  title = {{{OntoSoft}}: {{Capturing Scientific Software Metadata}}},
  shorttitle = {{{OntoSoft}}},
  booktitle = {Proceedings of the 8th {{International Conference}} on {{Knowledge Capture}}},
  author = {Gil, Yolanda and Ratnakar, Varun and Garijo, Daniel},
  year = {2015},
  month = oct,
  series = {K-{{CAP}} 2015},
  pages = {1--4},
  publisher = {Association for Computing Machinery},
  address = {New York, NY, USA},
  doi = {10.1145/2815833.2816955},
  isbn = {978-1-4503-3849-3}
}

@inproceedings{gil_OntoSoft_2016,
  title = {{{OntoSoft}}: {{A}} Distributed Semantic Registry for Scientific Software},
  shorttitle = {{{OntoSoft}}},
  booktitle = {2016 {{IEEE}} 12th {{International Conference}} on E-{{Science}} (e-{{Science}})},
  author = {Gil, Yolanda and Garijo, Daniel and Mishra, Saurabh and Ratnakar, Varun},
  year = {2016},
  month = oct,
  pages = {331--336},
  doi = {10.1109/eScience.2016.7870916}
}

@article{gruenpeter_D44_2023,
  title = {D4.4 - {{Guidelines}} for Recommended Metadata Standard for Research Software within {{EOSC}}},
  author = {Gruenpeter, Morane and Granger, Sabrina and Monteil, Alain and Chue Hong, Neil and Breitmoser, Elena and Antonioletti, Mario and Garijo, Daniel and Gonz{\'a}lez Guardia, Esteban and Gonzalez Beltran, Alejandra and Goble, Carole and {Soiland-Reyes}, Stian and Juty, Nick and Mejias, Gabriela},
  year = {2023},
  month = jun,
  doi = {10.5281/zenodo.10786147},
  langid = {english}
}

@article{hasselbring_MultiDimensional_2025,
  title = {Multi-{{Dimensional Research Software Categorization}}},
  author = {Hasselbring, Wilhelm and Druskat, Stephan and Bernoth, Jan and Betker, Philine and Felderer, Michael and Ferenz, Stephan and Hermann, Ben and Lamprecht, Anna-Lena and Linxweiler, Jan and Prat, Arnau and Rumpe, Bernhard and {Schoening-Stierand}, Katrin and Yang, Shinhyung},
  year = {2025},
  journal = {Computing in Science \& Engineering},
  pages = {1--10},
  issn = {1558-366X},
  doi = {10.1109/MCSE.2025.3555023}
}

@article{iglezakis_Modelling_2023,
  title = {Modelling {{Scientific Processes With}} the M4i {{Ontology}}},
  author = {Iglezakis, Dorothea and Terzijska, D{\v z}ulia and Arndt, Susanne and Leimer, Sophia and Hickmann, Johanna and Fuhrmans, Marc and Lanza, Giacomo},
  year = {2023},
  month = sep,
  journal = {Proceedings of the Conference on Research Data Infrastructure},
  volume = {1},
  issn = {2941-296X},
  doi = {10.52825/cordi.v1i.271},
  copyright = {Copyright (c) 2023 Dorothea Iglezakis, D{\v z}ulia Terzijska , Susanne Arndt, Sophia Leimer, Johanna Hickmann, Marc Fuhrmans, Giacomo Lanza},
  langid = {english}
}

@article{ison_BiotoolsSchema_2021,
  title = {{{BiotoolsSchema}}: {{A}} Formalized Schema for Bioinformatics Software Description},
  shorttitle = {{{BiotoolsSchema}}},
  author = {Ison, J. and Ienasescu, H. and Rydza, E. and Chmura, P. and Rapacki, K. and Gaignard, A. and Schw{\"a}mmle, V. and Van Helden, J. and Kala{\v s}, M. and M{\'e}nager, H.},
  year = {2021},
  journal = {GigaScience},
  volume = {10},
  number = {1},
  issn = {2047-217X},
  doi = {10.1093/gigascience/giaa157},
  langid = {english}
}

@misc{jones_CodeMeta_2023,
  title = {{{CodeMeta}}: An Exchange Schema for Software Metadata. {{Version}} 3.0.},
  author = {Jones, Matthew B. and Boettiger, Carl and Mayes, Abby Cabunoc and Smith, Arfon M. and Gruenpeter, Morane and Lorentz, Valentin and Morrell, Thomas and Garijo, Daniel and Slaughter, Peter and Niemeyer, Kyle E. and Gil, Yolanda and Fenner, Martin and Nowak, Krzysztof and Hahnel, Mark and Coy, Luke and Allen, Alice and Crosas, Merc{\`e} and Sands, Ashley and Chue Hong, Neil and Cruse, Patricia and Katz, Daniel S and Goble, Carole and Mecum, Bryce and {Gonzalez-Beltran}, Alejandra and Ross, Noam},
  year = 2023,
  publisher = {Github},
  url = {https://codemeta.github.io/terms/},
  urldate = {2025-10-28},
  langid = {english}
}

@article{kuckertz_DataDesc_2024,
  title = {{{DataDesc}}: {{A}} Framework for Creating and Sharing Technical Metadata for Research Software Interfaces},
  shorttitle = {{{DataDesc}}},
  author = {Kuckertz, Patrick and G{\"o}pfert, Jan and Karras, Oliver and Neuroth, David and Sch{\"o}nau, Julian and Pueblas, Rodrigo and Ferenz, Stephan and Engel, Felix and Pflugradt, Noah and Weinand, Jann M. and Nie{\ss}e, Astrid and Auer, S{\"o}ren and Stolten, Detlef},
  year = {2024},
  month = oct,
  journal = {Patterns},
  pages = {101064},
  issn = {2666-3899},
  doi = {10.1016/j.patter.2024.101064},
  copyright = {All rights reserved}
}

@inproceedings{palavitsinis_Evaluation_2009,
  title = {Evaluation of a {{Metadata Application Profile}} for {{Learning Resources}} on {{Organic Agriculture}}},
  booktitle = {Metadata and {{Semantic Research}}},
  author = {Palavitsinis, Nikos and Manouselis, Nikos and Sanchez Alonso, Salvador},
  editor = {Sartori, Fabio and Sicilia, Miguel Angel and Manouselis, Nikos},
  year = {2009},
  series = {Communications in {{Computer}} and {{Information Science}}},
  pages = {270--281},
  publisher = {Springer},
  address = {Berlin, Heidelberg},
  doi = {10.1007/978-3-642-04590-5_26},
  isbn = {978-3-642-04590-5},
  langid = {english}
}

@inproceedings{schwarz_Ontological_2019,
  title = {Ontological {{Integration}} of {{Semantics}} and {{Domain Knowledge}} in {{Energy Scenario Co-simulation}}},
  booktitle = {Proceedings of the 11th {{International Joint Conference}} on {{Knowledge Discovery}}, {{Knowledge Engineering}} and {{Knowledge Management}}},
  author = {Schwarz, Jan and Lehnhoff, Sebastian},
  year = {2019},
  pages = {127--136},
  publisher = {{SCITEPRESS - Science and Technology Publications}},
  address = {Vienna, Austria},
  doi = {10.5220/0008069801270136},
  isbn = {978-989-758-382-7},
  langid = {english}
}

@article{shin_Realtime_1994,
  title = {Real-Time Computing: A New Discipline of Computer Science and Engineering},
  shorttitle = {Real-Time Computing},
  author = {Shin, K.G. and Ramanathan, P.},
  year = {1994},
  month = jan,
  journal = {Proceedings of the IEEE},
  volume = {82},
  number = {1},
  pages = {6--24},
  issn = {1558-2256},
  doi = {10.1109/5.259423}
}

@article{venkatesh_Technology_2008,
  title = {Technology {{Acceptance Model}} 3 and a {{Research Agenda}} on {{Interventions}}},
  author = {Venkatesh, Viswanath and Bala, Hillol},
  year = {2008},
  month = may,
  journal = {Decision Sciences - DECISION SCI},
  volume = {39},
  pages = {273--315},
  doi = {10.1111/j.1540-5915.2008.00192.x}
}

@article{venkatesh_Theoretical_2000,
  title = {A {{Theoretical Extension}} of the {{Technology Acceptance Model}}: {{Four Longitudinal Field Studies}}},
  shorttitle = {A {{Theoretical Extension}} of the {{Technology Acceptance Model}}},
  author = {Venkatesh, Viswanath and Davis, Fred D.},
  year = {2000},
  month = feb,
  journal = {Management Science},
  volume = {46},
  number = {2},
  pages = {186--204},
  issn = {0025-1909},
  doi = {10.1287/mnsc.46.2.186.11926}
}

@article{venkatesh_User_2003,
  title = {User {{Acceptance}} of {{Information Technology}}: {{Toward}} a {{Unified View}}},
  shorttitle = {User {{Acceptance}} of {{Information Technology}}},
  author = {Venkatesh, Viswanath and Morris, Michael G. and Davis, Gordon B. and Davis, Fred D.},
  year = {2003},
  journal = {MIS Quarterly},
  volume = {27},
  number = {3},
  eprint = {30036540},
  eprinttype = {jstor},
  pages = {425--478},
  issn = {0276-7783},
  doi = {10.2307/30036540}
}

@book{zeng_Metadata_2022,
  title = {Metadata},
  author = {Zeng, Marcia Lei and Qin, Jian},
  year = {2022},
  edition = {Third edition},
  publisher = {Facet Publishing},
  address = {London},
  isbn = {978-1-78330-588-9}
}

@misc{ferenz_ERSmeta_2025a,
  title = {{{ERSmeta}}},
  author = {Ferenz, Stephan},
  year = 2025,
  month = oct,
  doi = {10.5281/zenodo.17465772},
  howpublished = {Zenodo}
}

@article{ferenz_Questionnaire_2025,
  title = {Questionnaire for {{Evaluating}} a {{Metadata Schema}}  for {{Energy Research Software}}},
  author = {Ferenz, Stephan and Werth, Oliver},
  year = 2025,
  month = oct,
  publisher = {Zenodo},
  doi = {10.5281/zenodo.17465939},
  urldate = {2025-10-29},
  langid = {english}
}

@misc{ferenz_ERSmeta_2025b,
  title = {{{ERSmeta Evaluation Scripts}}},
  author = {Ferenz, Stephan},
  year = 2025,
  month = dec,
  doi = {10.5281/zenodo.17936836},
  urldate = {2025-12-15},
  howpublished = {Zenodo}
}

@misc{hulk_Open_2025,
  title = {Open {{Energy Metadata}} ({{OEMetadata}})},
  author = {H{\"u}lk, Ludwig and Huber, Jonas and Hofmann, Christian and Muschner, Christoph},
  year = 2025,
  month = jan,
  url = {https://github.com/OpenEnergyPlatform/oemetadata},
  urldate = {2025-12-16},
  copyright = {MIT}
}

@misc{knublauch_Shapes_2017,
  title = {Shapes {{Constraint Language}} ({{SHACL}})},
  author = {Knublauch, Holger and Kontoskostas, Dimitris},
  year = 2017,
  month = jul,
  url = {https://www.w3.org/TR/shacl/},
  urldate = {2025-12-16},
  langid = {english}
}

@misc{ferenz_Dataset_2026,
  title = {Dataset of the {{Evaluation}} of a {{Metadata Schema}}  for {{Energy Research Software}}},
  author = {Ferenz, Stephan and Werth, Oliver},
  year = 2026,
  month = jan,
  publisher = {Zenodo},
  doi = {10.5281/zenodo.17935022},
  urldate = {2026-01-05},
  langid = {english}
}
